\documentclass[twocolumn, twocolappendix]{aastex631}

\newcommand{\Msun}{\ensuremath{\textrm{\,M}_{\odot}}}

\usepackage{amsmath}
\shorttitle{GAP I -- Ground-Based Asteroseismology of Massive Stars}
\shortauthors{Shitrit \& Arcavi}

\graphicspath{{./}}

\begin{document}

\title{The Global Asteroseismology Project Proof of Concept: Asteroseismology of Massive Stars with Continuous Ground-Based Observations}

\author[0000-0001-9511-6054]{Noi Shitrit}
\affiliation{The School of Physics and Astronomy, Tel Aviv University, Tel Aviv 69978, Israel}

\author[0000-0001-7090-4898]{Iair Arcavi}
\affiliation{The School of Physics and Astronomy, Tel Aviv University, Tel Aviv 69978, Israel}

\correspondingauthor{Noi Shitrit}
\email{noyshitrit@tauex.tau.ac.il}

\begin{abstract}
Massive ($\gtrsim8\Msun$) stars are the progenitors of many astrophysical systems, yet key aspects of their structure and evolution are poorly understood. Asteroseismology has the potential to solve these open puzzles, however, sampling both the short period pulsations and long period beat patterns of massive stars poses many observational challenges. Ground-based single-site observations require years or decades to discern the main oscillation modes. Multi-site campaigns were able to shorten this time span, but have not been able to scale up to population studies on samples of objects. Space-based observations can achieve both continuous sampling and observe large numbers of objects, however, most lack the multi-band data that is often necessary for mode identification and removing model degeneracies. Here, we develop and test a new ground-based observational strategy for discerning and identifying the main oscillation modes of a massive star in a few months, in a way that can be scaled to large samples. We do so using the Las Cumbres Observatory - a unique facility consisting of robotic, homogeneous telescopes operating as a global network, overcoming most of the challenges of previous multi-site efforts, but presenting new challenges which we tailor our strategy to address. This work serves as the proof of concept for the Global Asteroseismology Project, which aims to move massive star asteroseismology from single-objects to bulk studies, unleashing its full potential in constraining stellar structure and evolution models. This work also demonstrates the ability of the Las Cumbres Observatory to perform multi-site continuous observations for various science goals. 
\end{abstract}


\section{Introduction} \label{sec:intro}
Massive ($\gtrsim 8 \Msun$) stars, although rare, play a key role in many astrophysical processes. They have a relatively short and energetic life that often ends in a supernova explosion, driving chemical and mechanical feedback to their host galaxy while leaving behind a neutron star or black hole. However, key aspects in the structure and evolution of massive stars, such as the degree and profile of their internal chemical mixing, their metallicity, and their rotation profile, are poorly constrained observationally.

One way forward is through asteroseismology, the study of the oscillations of stars and their connection to the internal stellar structure. Pulsating early-type OB stars can be divided into several categories, \cite[e.g.][]{Bowman_rev, Aerts_review, Kurts_rev}. The two main classes with oscillations primarily driven by cyclic variation in opacity (the so-called $\kappa$-mechanism) are slowly pulsating B-type (SPB) stars and $\beta$~Cephei (hereafter, $\beta$~Cep) type stars. SPB stars are intermediate mass stars, with a mass range of about 3 to 8 $\Msun$, while $\beta$~Cep stars, on which we focus here, have masses in the range of about 8 to 25 $\Msun$.

$\beta$~Cep stars oscillate in low order $p$ (pressure) and $g$ (gravity) modes with periods between approximately 2 and 8 hours \citep{2005ApJS..158..193S, Aerts_book, Bowman_rev}. Some $\beta$~Cep stars were found to have high-order $g$-modes with periods longer than a day \citep{Handler_2006_betaCep, 2009ApJ...698L..56H}. In general, they produce both short-pulsation periods (of order hours to days) and long-term beats (of order months) arising from adjacent oscillation modes. Sampling these different time scales presents a significant observational challenge. For single-site campaigns, limited to night-time data collection, the observing time span required to discern the oscillation frequencies can be very long (e.g., 21 years for the $\beta$~Cep star HD129929, \citealt{Aerts2003}).

Space-based telescopes naturally allow continuous observations, which reduce the required time span substantially. Missions such as MOST \citep[Microvariability and Oscillations of Stars;][]{2003PASP..115.1023W}, CoRoT \citep[Convection, Rotation, and Planetary Transits;][]{2009A&A...506..411A}, BRITE \citep[BRIght Target Explorer;][]{2014PASP..126..573W}, Kepler \citep{2010ApJ...713L..79K}, K2 \citep{2014PASP..126..398H}, and TESS \citep[Transiting Exoplanet Survey Satellite;][]{Ricker_2014} produced high-cadence long time-base precise light curves which can be used for asteroseismology. Indeed, \cite{2021NatAs...5..715P}, for example, used Kepler light curves spanning $\sim$1500 days of 26 SPB stars to study their internal mixing profiles. Another example is \cite{Handler_2019}, who identified at least 34 excited modes for the $\beta$~Cep star PHL 346 from its TESS light curve. 

Unfortunately, space telescopes are expensive and are often designed to address particular requirements for a specific science mission. For example, most missions so far did not provide color information, which is important for mode identification (i.e. assigning each mode with its harmonic degree $l$ and azimuthal order $m$; see e.g. \citealt{Aerts_book} and \citealt{mode_iden_B}). This identification increases the extent of astrophysical information gained from the asteroseismic analysis as it can lift model degeneracies. 

For slowly rotating stars (rotating at $\lesssim$15\% of their critical rotation velocity; \citealt{Bowman_rev}), mode identification can be performed by using rotational splitting of the oscillation modes \citep[e.g.][]{Aerts_book}. However, $\beta$~Cep stars display a variety of rotation rates, with some having $v \sin\left(i\right)$ as high as a few hundred km\,s$^{-1}$ \citep{2012MNRAS.424.2380H}, where $i$ is the inclination angle of the rotation axis relative to the viewing angle. For such stars, mode identification can often only be done through multi-band measurements of their oscillations. Multi-band observations probe changes in both the temperature and geometrical cross-section of pulsating stars. These changes are correlated with the spherical harmonic of the oscillation mode. Therefore, comparing the ratios between amplitudes and the differences between phases in different bands to the ones predicted by models, can be used to identify the modes, also for fast rotators \citep[e.g.][]{2002A&A...392..151D, 2015MNRAS.446.1438D}.

Multi-site ground-based observations can overcome both the day and night interruptions that affect single-site ground-based campaigns (see \citealt{Aerts_review} and references therein), and can provide color information, not usually available in space-based campaigns, for mode identification. Indeed, \cite{1991ApJ...378..326W} obtained over 264 hours of continuous observations of a pulsating pre white dwarf star, identifying over 100 pulsation modes and determining its mass, rotation period, magnetic field, and constraining its compositional stratification, using the Whole Earth Telescope (WET) project \citep{1990ApJ...361..309N}. WET was a coordinated effort among existing telescopes worldwide to perform a few continuous observation campaigns per year of order 10-day durations. \cite{Handler_2006_betaCep} observed the $\beta$~Cep star 12 (DD) Lacertae (``12 Lac'') using a multi-site ground-based campaign, spanning 750 hours of observations over 190 nights, detected 23 sinusoidal signals, and performed mode identification from their multi-band observations (which allowed them to derive the spherical degree $l$ of the five strongest modes and constraints on $l$ for the weaker ones). The number of modes they detected was larger than predicted by standard models, which might be due to a gap in our understanding of the chemical structure of 12 Lac. 

These efforts demonstrate the potential of continuous ground-based observations. However, they required a complicated and challenging observational setup involving coordination between several groups of astronomers, observatories, and systems. In addition, differences between telescopes add challenges and uncertainties to the data analysis. Such methods, therefore, end up being difficult to scale up for studying large numbers of objects. While deep and impactful insights can be gained from single object studies, the full potential of asteroseismology in constraining stellar structure and evolution models lies in sample studies \citep[e.g.][]{2021NatAs...5..715P}. 

Here, we propose and test a new method with a new facility that we claim can be used to perform population sample studies of massive stars, and can be applied to a variety of other science cases. We propose to use the Las Cumbres Observatory \citep{LCO}, which differs from previously used global networks in that it is fully robotic and homogeneous (i.e. sets of identical telescopes, instruments, and filters are installed around the globe). While this facility has been operational for almost 10 years, it has yet to be used extensively for continuous observations in general, and for massive star asteroseismology in particular. A main challenge posed by Las Cumbres is its multi-purpose nature. Since this facility can not be dedicated to a single science program for extended periods of time, we propose to move from \textit{weeks-scale continuous observation runs}, used previously, to \textit{days-scale continuous observation runs} executed roughly once a week. In this work we show for the first time that such observations are technically possible with the Las Cumbres Observatory, that it provides the necessary photometric precision to perform asteroseismic analysis of massive stars, that all of this can be done in multiple bands simultaneously, and that such an observing strategy can reproduce single-site decade-long campaigns in a few weeks to months.

The Las Cumbres Observatory consists of ten 0.4-meter (0.4\,m) telescopes, fifteen 1-meter (1\,m) telescopes, and two 2-meter (2\,m) telescopes (all telescopes are equipped with standard imagers and multiple filters; Table~\ref{tab:cam_table}). The different telescopes of each size are identical, they are coordinated robotically, and are distributed around the globe in the United States (Texas, Hawaii), Chile, Spain (Tenerife), South Africa, Israel, and Australia.

\begin{deluxetable*}{@{\extracolsep}llllll@{}}
\tablecaption{Properties of the imagers used in this work.} \label{tab:cam_table}
\tablehead{\colhead{Camera name} & \colhead{Telescope class} & \colhead{Pixel scale\tablenotemark{a}} & \colhead{Field of View} & \colhead{Overhead per frame} & \colhead{ Filter options } \\ 
\colhead{} & \colhead{} & \colhead{(\arcsec/pixel)} & \colhead{} & \colhead{(s)} & \colhead{} } 
\startdata
Sinistro & 1\,m & 0.389 ($1 \times 1$ binning) & $26\arcmin\ \times 26\arcmin $ & 28 & 21 \\
SBIG & 1\,m & 0.464 ($2 \times 2$ binning) & $15.8\arcmin\ \times \ 15.8\arcmin$ & 15 & 21 \\
SBIG 6303 & 0.4\,m & 0.571 ($1 \times 1$ binning) & $29\arcmin\ \times \ 19\arcmin$ & 14 & 9 \\
\enddata
\tablenotetext{a}{After the stated standard binning.}
\tablecomments{The SBIG cameras are no longer available on the 1\,m telescopes.}
\end{deluxetable*}
From our experience with the Las Cumbres Observatory, its scheduling constraints due to multiple competing programs, and weather constraints at its different sites, we posit that 48-hour continuous observing blocks can be reasonably executed regularly. Thus, in Section \ref{sec:simulations} we propose a new observational strategy, using 48-hour blocks of observations over multiple weeks, and show, through simulations, that it is effective for massive-star oscillation mapping. Then, in Section \ref{sec:observations}, we collect actual Las Cumbres Observatory data from different sites and telescope classes for two massive stars in five observational campaigns, to demonstrate that the Las Cumbres Observatory can indeed perform 48-hour continuous runs in practice. Next, in Section \ref{sec:analysis} we construct an automatic analysis pipeline to handle the large number of images that such observations produce, exploring the best defocusing amount required for various Las Cumbres Observatory telescope classes, and the best image reduction parameters needed to achieve mili-magnitude photometric accuracy (as required for asteroseismic studies). Finally, in Section \ref{sec:lc_analysis} we test the quality of our observations and photometric extraction. We conclude and discuss our results in Section \ref{sec:conclusions}.

This paper is the first in a series describing the Global Asteroseismology Project (GAP) and its expected discoveries. The GAP was granted 9000 observing hours over three years (starting on 2023 August 1) on the 0.4m telescopes of the Las Cumbres Observatory network, on the basis of the proof of concept presented here. The time was granted as a key project, which guarantees its long-term status, and a high science priority to mitigate telescope oversubscription risks. Through GAP, we plan to collect multi-band high-cadence data of $ \sim 20 \; \beta$ Cep stars to complement single-band TESS data. This will allow us to perform complete seismological studies, including mode identification, of a sample of massive stars, almost tripling the current sample. The second paper in the series will detail our target selection process (Shitrit et al. in preparation).

\section{Targets}

Since our goal is to test a new observational strategy, we choose two $\beta$~Cep stars for which full asteroseismic analyses were already conducted (through both space-based and long-term ground-based observations): HD129929 \citep{Aerts2003} and HD180642 \citep{2009A&A...506..111D}, to demonstrate our proposed approach. 

\subsection{HD129929}

HD129929 is a B3V $\beta$~Cep star with a $V$-band magnitude of $8.1$ and an estimated mass of $10\Msun$ \citep{Hill1974_southern, Aerts2003}. The first to detect its variability was \cite{1981_HD129929_first_var_detec}, with a detailed study of the variability later performed by \cite{Waelkens_Rufener_1983}, identifying three main frequency modes. The more recent study by \cite{Aerts2003}, who observed HD129929 over a period of 21.2 years with the Geneva multi-color photometer, detected a beating phenomenon with at least six different frequencies, including two frequency multiplets of average spacing $\sim 0.0121$\,c\,day$^{-1}$ (cycles per day).

HD129929 is at right ascension $221.6073$ deg and declination $-37.2222$ deg and is therefore mainly visible from the southern hemisphere. We observed it from the Las Cumbres Observatory sites in Chile, South Africa, and Australia (there is a $\sim 25$-minute observational gap between Chile and Australia for this position on the sky).

During the first two HD129929 observing campaigns, no 0.4\,m telescopes were yet available on the network, and the 1\,m telescopes were being transitioned from SBIG to Sinistro cameras (Table~\ref{tab:cam_table}), with some telescopes still using SBIGs and some on Sinistros\footnote{Today the 1m network is completely homogeneous}. In the third campaign, 0.4\,m telescopes were available at all southern sites all of which were equipped with the same camera type (SBIG).

\subsection{HD180642}
HD180642 is a B1.5 II-III $\beta$~Cep star with a $V$-band magnitude of $8.29$ and an estimated mass in the range $11.4\Msun$ -- $11.8\Msun$ \citep{HD180642_2009}. \cite{HD180642_2009} combined observations of HD180642 from the CoRoT and Hipparcos space telescopes, and additional ground-based facilities across 18 years. They found 11 independent frequencies (and 22 three-resonance combinations) including the main frequency which was already detected in a previous study ($5.4871$\,c\,day$^{-1}$;  \citealt{Aerts_2000_HD180642}).

HD180642 is at right ascension $289.3117$ deg and declination $1.0594$ deg, making it visible from all Las Cumbres Observatory sites. However, due to a temporary focus mechanism problem with the 0.4\,m telescope in South Africa, and scheduling constraints, we observed HD180642 with the 0.4\,m telescopes in Chile, Australia, Tenerife, and Hawaii.

\section{Simulations} \label{sec:simulations}
We create an artificial light curve using:
\begin{eqnarray} \label{eq:artificial_signal}
F\left(t\right) = \sum_{i=0}^{N}a_i \cdot \sin\left(w_i\left(t-t_0\right) + \phi_i\right) \\\nonumber
w_i = 2\pi f_i,
\end{eqnarray}
where $F\left(t\right)$ is the flux at time $t$, $N$ is the number of excited modes, $a_i$, $f_i,$ and $\phi_i$ are their amplitudes, frequencies, and phases (respectively), and $t_0$ is a global time shift. We use the frequencies, amplitudes, and phases as determined by \cite{2004AA...415..241A} for HD129929 and the first 10 modes detected by \cite{HD180642_2009} for HD180642 in the $V$-band. For now, we set the time shift to $t_0=0$ (we will change this later for the Monte Carlo simulations in Section~\ref{sec:monte_carlo_sim}, and for comparison simulations in Section~\ref{sec:periodograms}).

\subsection{Single Simulations} \label{sec:single_sims}

As stated above, we simulate observation campaigns of weekly ``runs'' of 48 hours each. We denote the total time span of the observations $X$, the number of hours observed continuously per campaign $Y$, and the sampling cadence (i.e., the time between consecutive images taken within each run) $Z$. The simulations include a randomized $\pm2$-day offset at the start of each run (which is expected due to weather and scheduling constraints and can help remove aliases) and a randomly timed weather gap of two hours in every 48-hour run.  We add random Gaussian-distributed noise with a standard deviation of $3$ parts per thousand (PPT) to the flux values in all of our simulations. This is a typical relative photometry uncertainty value for Las Cumbres 1\,m telescopes\footnote{The 0.4\,m telescope cameras are currently being replaced, but we verified that increasing the noise by up to an order of magnitude does not significantly affect our results.}.

We extract the oscillation frequencies of the simulated data using a prewhitening process (see \citealt{Aerts_book} for full details)\footnote{We use the iterative prewhitening code of the IvS package - \url{https://github.com/IvS-KULeuven/IvSPythonRepository}}. We continue the prewhitening process as long as: (1) the signal to noise ratio (SNR) of the amplitude of the strongest mode is $>$4 in the periodogram (following \citealt{1993AA...271..482B}), and (2) the ratio of the amplitude of the strongest mode to that from the previous step is $>$1/3. The first condition is a significance threshold and the second condition is a stopping criterion introduced to avoid false modes which also cross the significance threshold\footnote{This stopping criterion is based on the two stars studied here. A more extensive study of stopping conditions is beyond the scope of this paper since the GAP relies on recovering known modes identified by other sources.}. Examples of the prewhitening process for the different targets can be seen in Figures~\ref{fig:prewhitening_HD129929} and \ref{fig:prewhitening_HD180642} in Appendix~\ref{appendix:prewhitening}.

\begin{figure}
\includegraphics[width=1\linewidth]{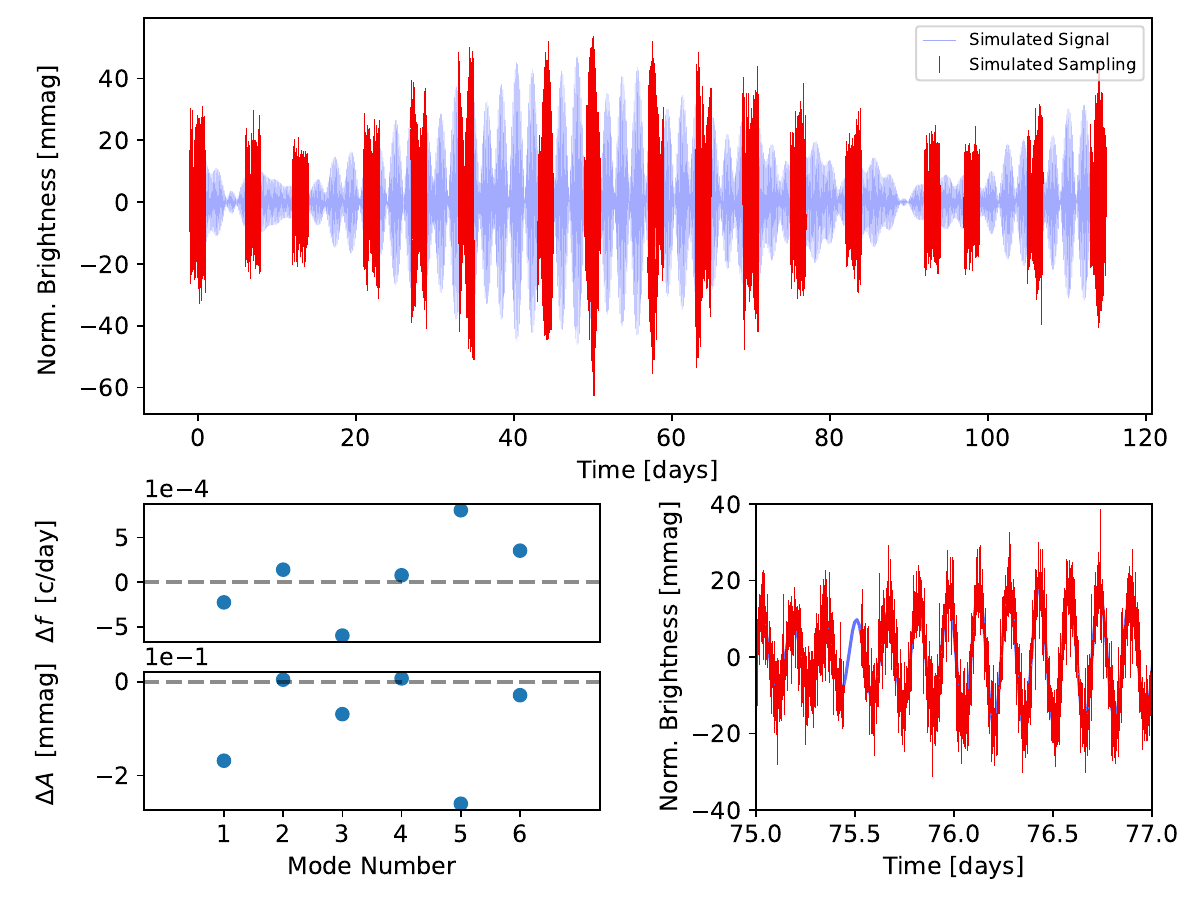}
\caption{Sequences of weekly 48-hour observation runs can recover the oscillation frequencies of the massive star HD129929, previously obtained during two decades of single-site observations, in 17 weeks. Top: Simulated signal of HD129929 (blue line; based on the modes detected by \citealt{2004AA...415..241A}) and simulated sampling of it during 17 weekly 48-hour runs of 60-second cadence per run (red points; with a measurement uncertainty of 3\,PPT). The simulations include $\pm2$-day randomness at the start of each 48-hour run (which is expected due to weather and scheduling constraints and can help remove aliases), and a randomly timed two-hour weather gap in each 48-hour run, which can be seen in the bottom right panel, displaying one such run. 
Bottom left: The difference between the frequencies ($\Delta f$) and amplitudes ($\Delta A$) recovered from our simulated data, using a prewhitening process (see Figure~\ref{fig:prewhitening_HD129929} in Appendix~\ref{appendix:prewhitening} for more details) and the ``true'' frequencies and amplitudes input to the simulation.
The input frequencies and amplitudes are recovered to within $\sim 10 ^{-4}$\,c\,day$^{-1}$ and $\sim 10^{-1}$\,mmag respectively. The prewhitening process used to extract the frequencies is presented in Figure \ref{fig:prewhitening_HD129929} in Appendix~\ref{appendix:prewhitening}.}
\label{fig:pre_observation_sim_HD129929}
\end{figure}

\begin{figure}
\includegraphics[width=1\linewidth]{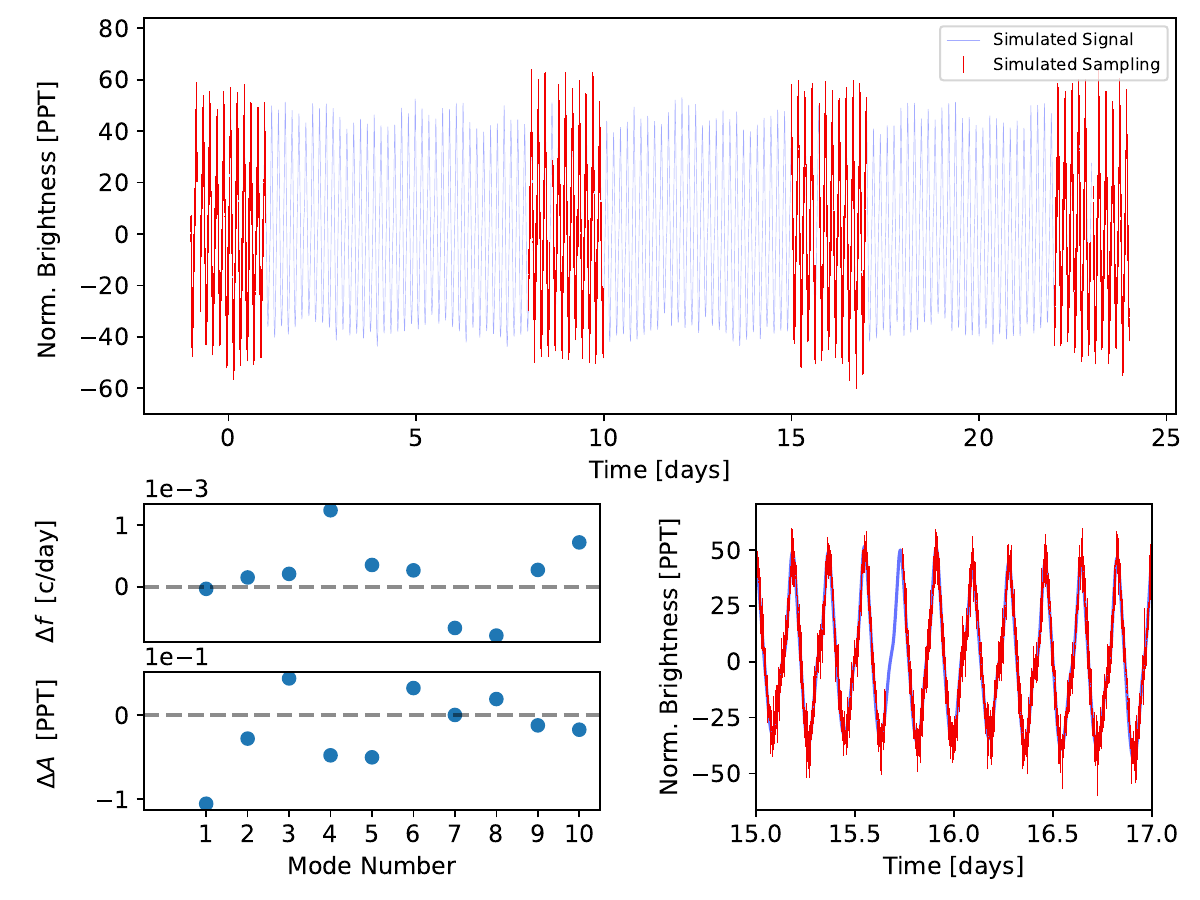}
\caption{Same as Figure~\ref{fig:pre_observation_sim_HD129929}, but for the massive star HD180642. Here, only 4 sequences of weekly 48-hour consecutive observations recover the top 10 oscillation frequencies and amplitudes obtained through 18.8 years of single-site observations to within $\sim 10 ^{-3}$\,c\,day$^{-1}$ and $\sim 10^{-1}$\,PPT respectively. The prewhitening process used to extract the frequencies is presented in Figure \ref{fig:prewhitening_HD180642} in Appendix~\ref{appendix:prewhitening}.}
\label{fig:pre_observation_sim_HD180642}
\end{figure}

The results of the simulations are presented in Figures~\ref{fig:pre_observation_sim_HD129929} and \ref{fig:pre_observation_sim_HD180642}, and in Table \ref{tab:values_HD129929_HD180642}. We find that 17 weekly 48-hour, 60-second cadence (i.e., $X = 17 \;$weeks, $Y ~= ~48 \;$hours, and $Z = 60 \;$seconds) runs can recover the full frequency spectrum of HD129929, as observed by \cite{Aerts2003} in 21 years, here in under four months, to an accuracy of $\sim 10 ^{-4}$\,c\,day$^{-1}$ in frequency and $\sim 10^{-1}$\,mmag in amplitude. For HD180642 (Figure~\ref{fig:pre_observation_sim_HD180642}), only 4 weeks (i.e., $X = 4 \;$weeks) of 48-hour observation blocks are required to recover the ten highest amplitude modes from the 18.8 year-long data set of \cite{HD180642_2009} to within $\sim 10 ^{-3}$\,c\,day$^{-1}$ in frequency and $\sim 10^{-1}$\,PPT in amplitude. 

The difference between the $X$ values required for each target is due mainly to differences in the spacing between their oscillation frequencies. HD129929 has an average frequency spacing of $\sim$ 0.012\,c\,day$^{-1}$, for all detected frequencies \citep{Aerts2003}, while HD180642 has an average frequency spacing of $\sim$1.24\,c\,day$^{-1}$ for frequencies between 4.32\,c\,day$^{-1}$ and 25.9\,c\,day$^{-1}$ \citep{refId0}. These two stars are roughly at opposite extremes of the GAP sample in terms of their frequency spacing (Shitrit et al. in preparation), and thus represent the boundaries of the $X$ values that will be required for the various stars in the sample.

\begin{deluxetable*}{CCCCCC}
\tablecaption{Input and recovered frequencies ($f$) and amplitudes ($A$) for our test targets. The input values are taken in the $V$ band from \citealt{2004AA...415..241A} for HD129929 with amplitudes in mmag, and from  \citealt{HD180642_2009} for HD180642 with amplitudes in PPT).} \label{tab:values_HD129929_HD180642}
\tablehead{\colhead{Mode} & \colhead{$f_{in}$} & \colhead{$f_{rec}$} & \colhead{$A_{in}$} & \colhead{$A_{rec}$} & \colhead{SNR}\\
\colhead{No.} & \colhead{(c\,day$^{-1}$)} & \colhead{(c\,day$^{-1}$)} & \colhead{(mmag/PPT)} & \colhead{(mmag/PPT)} & \colhead{}} 
\startdata
\multicolumn{5}{c}{\textrm{HD129929}} \\
\hline
1 & 6.461699 & 6.461472 \pm 0.000057 & 11.8 & 11.631665 \pm 0.140547 & 36.24\\
2 & 6.978305 & 6.978445 \pm 0.000054 & 10.3 & 10.304667 \pm 0.116364 & 37.13\\
3 & 6.449590 & 6.448991 \pm 0.000119 & 9.1 & 9.031067 \pm 0.226550 & 31.99\\
4 & 6.990431 & 6.990508 \pm 0.000041 & 7.5 & 7.507319 \pm 0.065104 & 27.61 \\
5 & 6.590940 & 6.591746 \pm 0.000022 & 4.9 & 4.639817 \pm 0.021042 & 17.33\\
6 & 6.966172 & 6.966525 \pm 0.000049 & 4.8 & 4.771511 \pm 0.048805 & 14.19\\
\hline
\multicolumn{5}{c}{\textrm{HD180642}} \\
\hline
1 & 5.486889 & 5.486855 \pm 0.000113 & 36.96 & 36.854273 \pm 0.188457 & 704.93\\
2 & 10.97374 & 10.973890 \pm 0.000436 & 6.31 & 6.281827 \pm 0.124152 & 120.23\\
3 & 16.460665 & 16.460873 \pm 0.000506 & 4.05 & 4.093468 \pm 0.093844 & 77.03\\
4 & 0.299171 & 0.300408 \pm 0.000533 & 3.13 & 3.081897 \pm 0.074501 & 58.48 \\
5 & 6.324816 & 6.325168 \pm 0.000632 & 2.24 & 2.189547 \pm 0.062795 & 41.85\\
6 & 8.409185 & 8.409450 \pm 0.000779 & 1.56 & 1.591985 \pm 0.056254 & 32.51\\
7 & 7.254757 & 7.254090 \pm 0.000777 & 1.45 & 1.449963 \pm 0.051058 & 28.60\\
8 & 21.94756 & 21.946770 \pm 0.000802 & 1.27 & 1.288915 \pm 0.046848 & 24.83\\
9 & 11.81164 & 11.811912 \pm 0.000808 & 1.19 & 1.177548 \pm 0.043121 & 22.17\\
10 & 6.143358 & 6.144075 \pm 0.000906 & 1 & 0.982420 \pm 0.040373 & 16.58\\
\enddata
\tablecomments{The ${in}$ subscript stands for ``input'', and the ${rec}$ subscript stands for ``recovered''. The recovered values are for one of the realizations performed for each target, shown in Figure~\ref{fig:pre_observation_sim_HD129929} for HD129929 and in Figure~\ref{fig:pre_observation_sim_HD180642} for HD180642. The SNR is calculated as in \cite{1993AA...271..482B}, for HD129929 between 6 and 7 c\, day$^{-1}$ and for HD180642 between 0 and 20 c\,day$^{-1}$.}
\end{deluxetable*}

\subsection{Monte Carlo Simulations} \label{sec:monte_carlo_sim}

There could be combinations of $X$, $Y$, and $Z$ values other than the ones presented in Section~\ref{sec:single_sims} that can produce equally good results. In addition, the random timing of the 2-hour weather gap, the randomized $\pm 2$-day offset at the start of each run and the timing of the start of the campaign (denoted by $t_0$ in Equation~\ref{eq:artificial_signal}), can also influence the results. We do not conduct a full phase space study here, but we run four sets of 1000 simulations for HD129929 to check the effect of changing some of these parameters. In each simulation we choose a different realization for each of the randomized parameters, and repeat each set of 1000 simulations with a different value for $Y$ (the length of the continuous observation block): 12, 24, 36, and 48 hours. We then perform the prewhitening process for each of the simulated light curves to extract its frequencies and amplitudes. A summary of the results is presented in Figure~\ref{fig:HD129929_1000_sim_freq_amp} and in Table~\ref{tab:sim_1000_HD129929} in Appendix~\ref{appendix:MC_rand}. Figure~\ref{fig:HD129929_1000_sim_freq_amp} shows the differences between the recovered and input frequencies and amplitudes for the first, third and sixth modes (ordered by decreasing amplitude). The recovered frequencies and amplitudes are consistent with the input ones within the scatter of the simulations. As expected, the stronger modes are recovered more accurately and more consistently compared to the weaker ones. The quality of the results declines with decreasing observing run length ($Y$ value), also as expected.

\begin{figure}
\includegraphics[width=1\linewidth]{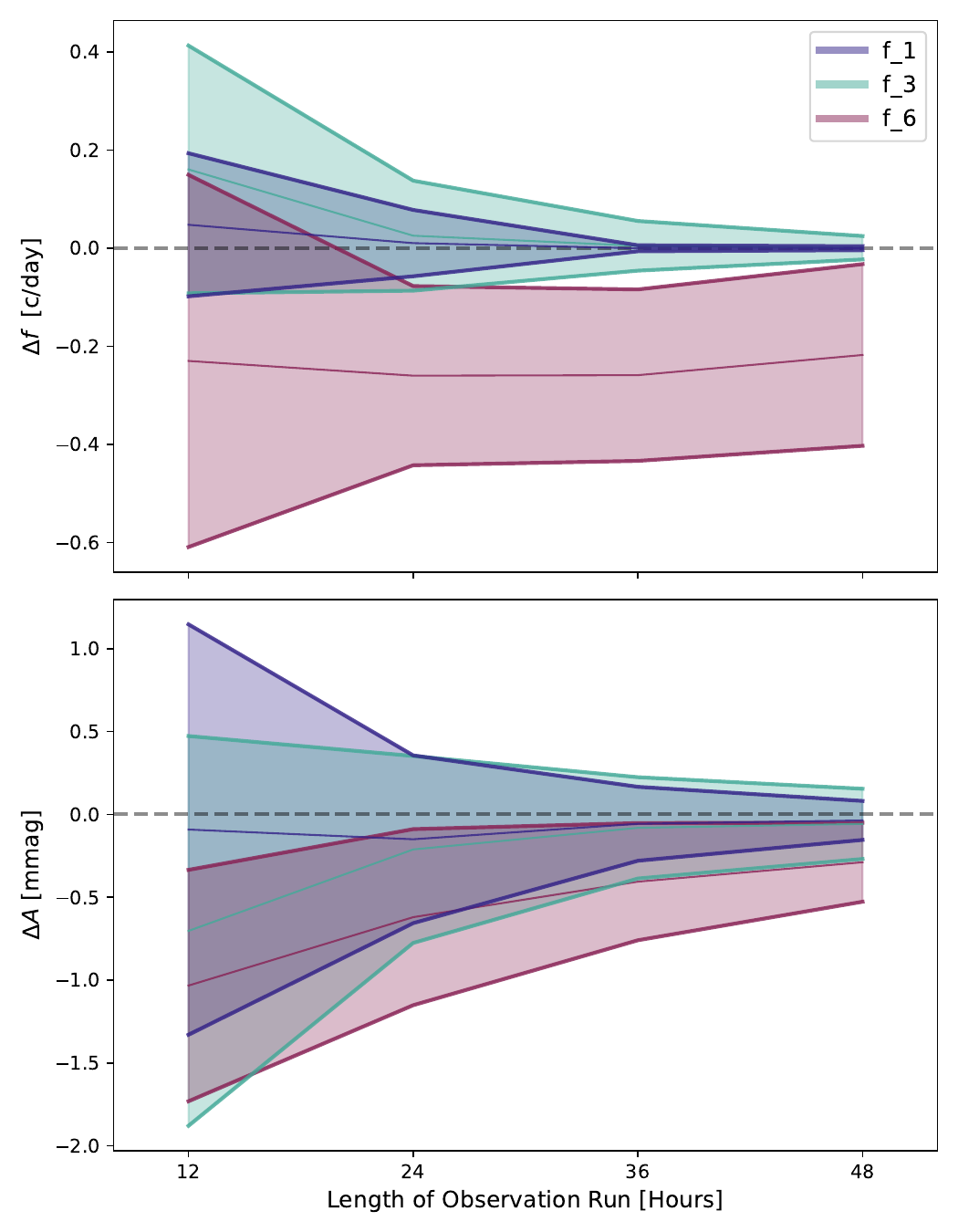}
\caption{Differences between recovered and input frequencies (top) and amplitudes (bottom) in 1000 Monte Carlo simulations for each of 4 different observing run lengths for HD129929. The average value $\pm1\sigma$ scatter of this difference from each set of 1000 simulations is shown for three out of the six observing modes (ordered in descending amplitude) for clarity. The recovered frequencies are consistent with the input ones to within the scatter of the simulations. Consistency and accuracy improve with increasing observing run length and mode amplitude, as expected.}
\label{fig:HD129929_1000_sim_freq_amp}
\end{figure}

We conclude that the results presented in Section \ref{sec:single_sims} are robust and that observation runs shorter than 48 hours degrade the accuracy of the recovered frequencies and amplitudes for HD129929. However, each star is different and might require different observation run lengths, depending on the desired mode recovery accuracy. Simulations such as these can be used to estimate the uncertainty in oscillation mode recovery for various observing run lengths and expected oscillation spectra.

\section{Observations} \label{sec:observations}

Having found that 48-hour runs can be used as building blocks for an observation sequence capable of reproducing the oscillation spectrum of our example targets, in principle, we turn to test this idea in practice. Namely, we test whether 48-hour runs can indeed be scheduled and executed by the Las Cumbres network, and whether the resulting images provide the necessary photometric accuracy to discern the oscillations. To do this, we conduct five observational campaigns (in total) of our two targets. 

Four of these campaigns, which we number 0--3, are for HD129929. Campaign 0 was a short $\sim$13-hour test across two sites on the 1\,m telescopes. Campaign 1 was the full $\sim$48-hour test across all three southern sites, also on the 1\,m telescopes. Campaign 2 was a $\sim$24-hour test on the 0.4\,m telescopes. Campaign 3 was a multi-band observation test on the 1\,m telescopes. As mentioned, multi-band observations are important for mode identification. We wish to test this setup because switching between filters resets the telescope guiding and can thus affect the photometric precision. We cycled between four filters: $B$, $V$, $R$, and $I$, with three consecutive images taken per filter. The final campaign, Campaign 4, was a $\sim$48-hour test for our second target, HD180642, using the 0.4\,m telescopes. 

\begin{figure}[]
\centering
\includegraphics[width=1\linewidth]{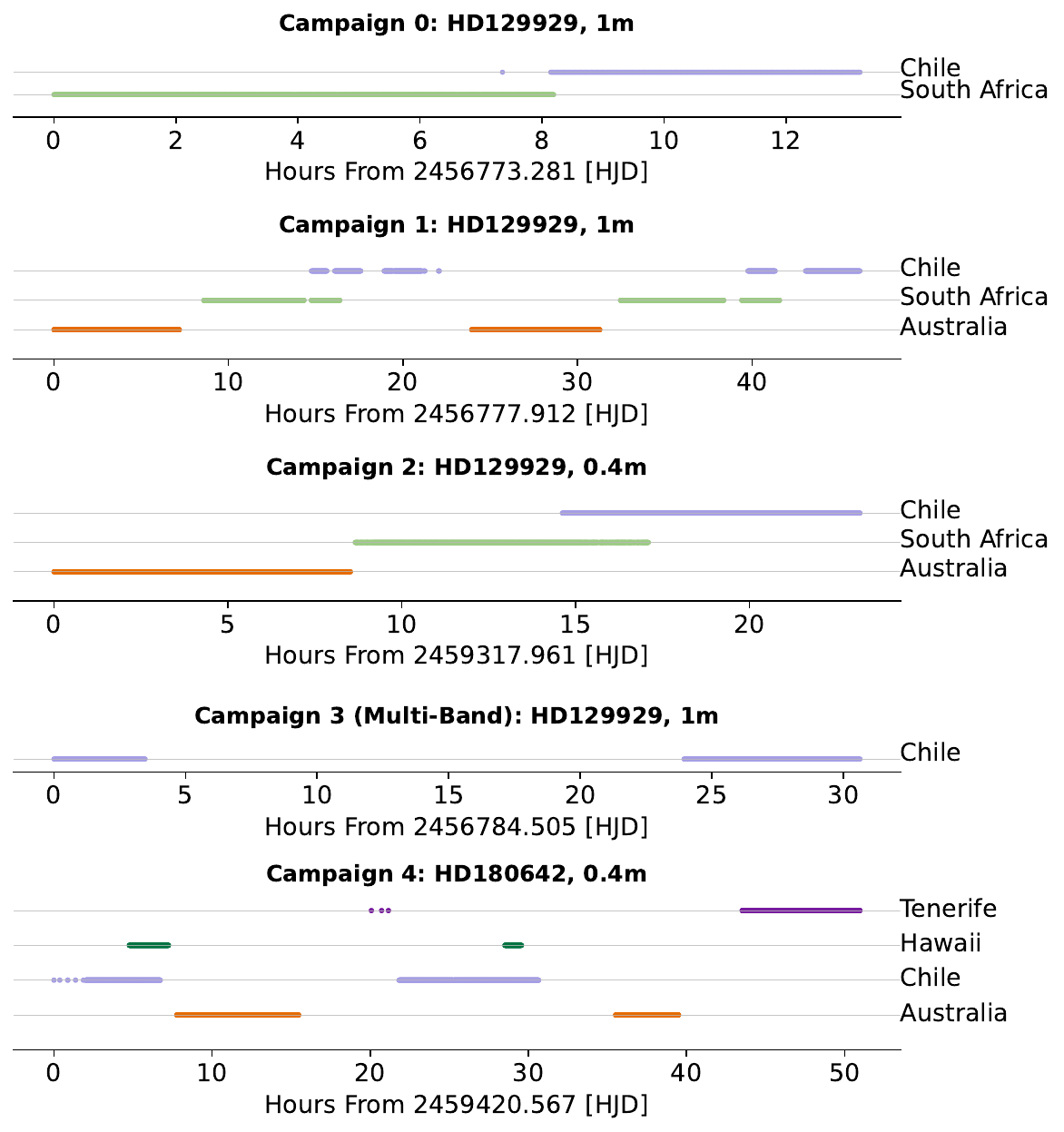}
\caption{Timelines of our five observational campaigns. Colored bands denote when observations were obtained from each site.}
\centering
\label{Fig:time_display}
\end{figure}

In total, we test two classes of telescopes (0.4\,m and 1\,m)\footnote{Since there are only two 2\,m telescopes, they are not suitable for continuous coverage.}, for two different targets in single and multi-band campaigns. Table~\ref{tab:obssummary} and Figure~\ref{Fig:time_display} summarize our observational campaigns. 

\begin{deluxetable*}{lllllllllp{3.3cm}}
\tablecaption{Summary of the five observation campaigns.} \label{tab:obssummary}
\tablehead{\colhead{Campaign} & \colhead{Target} & \colhead{Telescope} & \colhead{Defocus} & \colhead{Exp. } & \colhead{Filter} & \colhead{Time span} & \colhead{No. of } & \colhead{Pipeline} & \colhead{Dates} \\ 
\colhead{No.} & \colhead{} & \colhead{} & \colhead{(mm)} & \colhead{time (s)} & \colhead{} & \colhead{(hours)} & \colhead{images} & \colhead{} & \colhead{}}
\startdata
0 & HD129929 & 1\,m & 4 & 10 & \textit{V} & 13.2 & 1146 & \textsc{ORAC-DR} & 2014/04/25, 18:37:01 - 2014/04/26, 07:49:42 \\
1 & HD129929 & 1\,m & 4 & 10 & \textit{V} & 46.18 & 4732 & \textsc{ORAC-DR} & 2014/04/30, 09:45:32 - 2014/05/02, 07:56:42 \\
2 & HD129929 & 0.4\,m & 1 & 10 & \textit{V} & 23.16 & 3636 & \textsc{BANZAI} & 2021/04/13, 10:56:25 - 2021/04/14, 10:06:51 \\
3 & HD129929 & 1\,m & 4 & 10 & \textit{BVRI} & 10.14 & 586 & \textsc{ORAC-DR} & 2014/05/06, 23:59:09 - 2014/05/07, 03:27:09 \& 2014/05/07, 23:55:36 - 2014/05/08, 06:36:42\\
\hline
4 & HD180642 & 0.4\,m & 0 & 10 & \textit{V} & 50.96 & 5273 & \textsc{BANZAI} & 2021/07/25, 01:29:17 - 2021/07/27, 04:27:32 \\
\enddata
\tablecomments{All times UT.}
\end{deluxetable*}

Our targets are bright and can saturate the detectors (depending on the exposure time, airmass, and observing conditions, of course). Short exposures could solve the saturation issue, but are inefficient given the readout times of the imagers (Table~\ref{tab:cam_table}), and would push fainter reference stars to below the detection threshold. Instead, we choose to defocus the telescopes. Defocusing, however, creates non-trivial point-spread functions (PSFs). This presents some challenges in analyzing the data (see Section~\ref{sec:analysis}).

The amount of defocusing should be large enough to prevent saturation of the detector but not too large in order to minimize PSF distortions and avoid loss of stellar signal to background sky and noise. Before each campaign, we performed exposure tests to determine the appropriate defocusing amount for each telescope class and target. We used 10-second exposures per image and took 3 images per defocusing value. We then analyzed the images to find the one with the lowest defocusing amount and stellar counts still in the linear regime of the detector ($\lesssim55,000$ counts). We found that HD129929 required different amounts of defocusing for each telescope class while HD180642 did not require any defocusing (see Table~\ref{tab:obssummary}).

Images were initially processed by Las Cumbres Observatory. The pipeline that was in use during campaigns 0, 1, and 3 was based on the Observatory Reduction and Acquisition Control Data Reduction \citep[\textsc{ORAC-DR};][]{ORACDR2008, ORACDR2015} framework, which performs standard image processing, including bad-pixel masking, bias and dark subtraction, flat field correction, astrometric solution, source catalog production (including source identification and photometric extraction), zeropoint determination, and per-object airmass and barycentric time correction computation. Since the 2016A semester, Las Cumbres updated the standard pipeline to ``Beautiful Algorithms to Normalize Zillions of Astronomical Images'' \citep[\textsc{BANZAI};][]{BANZAI}, which was used for campaigns 2 and 4. \textsc{BANZAI} performs the same operations as ORAC-DR, with the exception of the barycentric time correction, and includes background subtraction from the overscan region \citep{LCO_two_years}. In addition, it uses a different algorithm for the astrometric solution \citep[Astrometry.net;][]{Astrometry}, and for source extraction \citep[\textsc{Source Extractor, hereafter SE};][]{SExtractor} and is \textsc{Python}-based (whereas \textsc{ORAC-DR} is written in \textsc{Perl}). 

\section{Producing the Light Curves}
\label{sec:analysis}

High cadence long-term observations like ours, result in thousands of images. Therefore, we construct a general automatic data reduction pipeline to produce relative-photometry light curves from a stack of defocused images.

As mentioned, defocusing the telescopes creates a non-trivial PSF. Instead of the ``natural'' Gaussian-like PSF that is expected from observing a point-like source, the defocused PSF is doughnut shaped, due to the shape of the telescope's primary mirror. Most of the standard programs for analyzing astronomical images assume a Gaussian PSF. Therefore, defocused images require a few adjustments to the default settings of these programs. 

\subsection{Cosmic Ray Removal}
\label{sec:cosmic_ray}
Cosmic rays are a significant contaminant for the astrometric solution (Section \ref{sec:Astrometry}). For example, for HD129929, for campaigns 0, 1, and 2, without cosmic ray removal, only $76.22\%$ of the images were astrometrically solved, while cosmic ray removal increased this percentage to $98.98\%$. Cosmic rays also produce false positives when performing source detection and extraction (Section \ref{sec:source_extract}).

To remove cosmic rays, we use the \texttt{detect\_cosmics()} function from the Speedy Cosmic Ray Annihilation Package in Python \citep[\textsc{Astro-SCRAPPY};][]{curtis_mccully_2018_1482019}, which is based on the Laplacian Cosmic Ray Identification algorithm \citep[\textsc{LA Cosmic};][]{van_Dokkum_2001}. We used the default \textsc{Astro-SCRAPPY} parameters except for \texttt{sigclip} which we set to $4.0$ and \texttt{objlim} which we set to $1000.0$ (see the \textsc{Astro-SCRAPPY} documentation\footnote{
\url{https://github.com/astropy/astroscrappy}} for a full description of the parameters).

\subsection{Astrometric Solution}
\label{sec:Astrometry}
We use a local installation of the \textsc{Astrometry.net} \citep{Astrometry} engine, and run the \texttt{solve-field} command on each image. We use \textsc{Astrometry.net}'s built-in downsampling to deal with the doughnut shaped PSF's caused by the defocusing. When using the downsampling option, \textsc{Astrometry.net} bins the images by $n{\times}n$ pixels before performing the source detection (with $n$ provided by the user). After finding the best solution, \textsc{Astrometry.net} scales the solution back to the original image. We find this process to be crucial for solving defocused images. It helps smooth the dark-centered regions of the sources, making source recognition by \textsc{Astrometry.net} more robust and efficient. In principle, downsampling is also useful for removing cosmic rays, however in our case it was not enough for most images, hence our use of \textsc{Astro-SCRAPPY} (see Section \ref{sec:cosmic_ray}).  We used a binning value of $n=9$ for HD129929. Since we do not perform any defocusing for HD180642, no downsampling was needed in that case.

\subsection{Source Extraction}
\label{sec:source_extract}
After obtaining the astrometric solution for the images, we perform aperture photometry on all sources in the image using \textsc{SE}. \textsc{SE} has a very detailed configuration file and a parameters file for specifying the output. We use a few different configuration files for the different campaigns (see Table~\ref{tab:SE_config} in Appendix~\ref{appendix:SExtractor}). Among the differences between the configuration files are the pixel scale of the different cameras, the convolution filter, and the aperture diameter.

The \textsc{SE} algorithm assumes a Gaussian PSF for source detection. Therefore, the doughnut shaped PSF (in the defocused images of HD129929) is sometimes detected by \textsc{SE} as two different Gaussian PSFs, and as a result, as two different sources. To avoid this, we perform a convolution with a top hat filter (using a $5 \times 5$ kernel) on the images. This step smooths the double-peaked PSF and removes many of the fictitious double sources.

To confirm that the relative number of counts between sources (for performing relative photometry) is conserved by the top-hat convolution, we create artificial images with two isolated sources (representing a target star and a reference star) and process them with \textsc{SE}, once with the top-hat convolution and once without. We then compare the ratio of counts between the sources in each case. We repeat the test for four different ratios of source fluxes. We find that the top-hat convolution conserves the relative counts consistently across the various flux ratios tested (Table~\ref{tab:tophat}) up to 2--4 parts per million (such differences have no notable effect on our analysis).

\begin{deluxetable}{DD}
\tablecaption{Conservation of relative flux with top-hat convolution for various input flux ratios (between two artificial sources).} \label{tab:tophat}
\tablehead{\multicolumn2c{Flux ratio}\hspace{2.6cm} & \multicolumn2c{Conservation ratio\tablenotemark{a}}}
\decimals
\startdata
1.2 \hspace{1.3cm}& 99.9994\% \\
1.5 & 100.0002\% \\
2 & 99.9994\% \\
3 & 99.9994\% \\
\enddata
\tablenotetext{a}{Ratio between the relative flux measured with no convolution to that with convolution.}
\end{deluxetable}

For HD180642, which we observed with no defocusing, we perform a convolution with a Gaussian filter using a $7 \times 7$ kernel.

We choose the best aperture size to use for each observing site and camera type by visually inspecting a selection of random images from that site and camera type.

The result of this stage is a catalog of sources with positions and instrumental aperture fluxes (aperture integrated, sky subtracted, total counts) per image. 

\subsection{Reference Stars}
\label{sec:reference_stars}
Performing relative photometry requires identifying a set of reference stars in all images. Having several reference stars helps reduce statistical uncertainties (and also systematic ones, in case one or some of the chosen reference stars are variable). However, massive stars are typically the brightest stars in the field of view, and there is usually only one more star detected at a similar signal to noise, to serve as a reference star. This is the case here, and we show that one reference star is enough to achieve the desired photometric precision to reveal the oscillations of our targets. 

Stars are associated across images by their celestial coordinates provided from the astrometric solution (Section \ref{sec:Astrometry}). However, due to the defocused PSFs, the astrometric solutions and the source extraction centroids are sometimes shifted between images. We thus associate stars across images based on their celestial separation between images being smaller than a pre-defined threshold of 5.4\arcsec\ for campaigns 0, 1, and 3, 2.34\arcsec\ for Campaign 2, and 1\arcsec\ for Campaign 4. These values are much smaller than the distance to neighboring sources, thus avoiding any source confusion.

For HD129929, the reference star used is CD-36 9599 which appeared in all of the images (except one, which we omit). For HD180642, two candidate reference stars appeared in all images. The brighter of the two, HD180662, is a multiple system \citep{ref_double_star}. We therefore choose the second brightest candidate as the reference star, HD180533. The details of the reference stars are presented in Table~\ref{tab:ref_table}. There is no indication of variability for these stars. We tested the constancy of each reference star using a comparison star (see Table~\ref{tab:ref_table}). For this test, we chose the nights that displayed the most stable conditions (for HD129929, from Campaign 0 in Chile and for HD180642, from Campaign 4 in Hawaii). We find that the reference star CD-36 9599 shows constant flux to within 2.53\,PPT, while the reference star HD180533 shows constant flux to within 13.73\,PPT on the nights tested. These levels are of order a few percent of the oscillations measured for our target stars (see below), and hence we consider the reference star luminosities to be constant.

\begin{deluxetable}{lcc}
\tablecaption{Our targets together with their respective reference and comparison stars.
} \label{tab:ref_table}
\tablehead{\colhead{Target} \hspace{55pt} & \colhead{HD129929} & \colhead{HD180642}}
\decimals
\startdata
RA (deg) & 221.6073 & 289.3117\\
DEC (deg) & -37.2222 & 1.0594\\
Spectral type & B3V & B1.5 II-III C\\
$V$ mag & 8.1 & 8.29 \\
\hline
Reference star & CD-36 9599 & HD180533\\
RA (deg) & 221.5375 & 289.2167\\
DEC (deg) & -37.2333 & 0.9992\\
Spectral type & K5 E & G6V C\\
$V$ mag & 9.43 & 9.96\\
Variance & 2.53 & 13.73\\
\hline
Comparison star & TYC 7309-640-1 & BD$+$00 4162\\
RA (deg) & 221.5940 & 289.4097\\
DEC (deg) & -37.1218 & 0.1963\\
$V$ mag & 10.11 & 10.51\\
\enddata
\end{deluxetable}

\subsection{Relative Photometry} \label{sec:relative_photometry}

The relative flux between our target and the reference star in each image is simply: 
\begin{equation} \label{eq:f_ratio}
    f_\mathrm{rel} = \frac{f_\mathrm{target}}{f_\mathrm{ref}},
\end{equation}
with the uncertainty calculated as:
\begin{equation} \label{eq:f_error}
    \Delta f_\mathrm
{rel} = \sqrt{\left(\frac{\Delta f_\mathrm{target}}{f_\mathrm{ref}}\right)^2+\left(\frac{f_\mathrm{target}\cdot \Delta f_\mathrm{ref}}{f_\mathrm{ref}^2}\right)^2}.
\end{equation}

Here, $f_\mathrm{rel}$ is the relative flux, and $f_\mathrm{target}$ and $f_\mathrm{ref}$ are the aperture flux of the target star and of the reference star (respectively). The $\Delta$ values are the uncertainties of each flux measurement from SE considering only photon and detector noise:
\begin{equation} \label{eq:fluxerr}
    \Delta f_\mathrm{target} = \Delta f_\mathrm{ref} = \sqrt{\Sigma_{i \in A}\left(\sigma_i^2 + \frac{p_i}{g_i}\right)},
\end{equation}
where $A$ is the set of pixels inside an aperture, $\sigma_i$ is the standard deviation of the noise estimated from the local background by \textsc{SE}, $p_i$ is the background-subtracted counts and $g_i$ is the effective detector gain, all per pixel $i$ in the aperture (see the \textsc{SE} documentation\footnote{\url{https://github.com/astromatic/sextractor}} for more details). 

\subsection{Removing False Measurements} \label{sec:filtering}

As a first filtering step of bad images, we remove all images in which:

\begin{itemize}
    \item The maximum counts of the star are over 50,000 or the minimum counts of the reference star are below 1,000 (including the background), to avoid images in which the target is too bright (in the non-linear regime of the detector) and the reference star is too faint (and suffers from low signal to noise).
    \item The error of the flux ratio (Equation~\ref{eq:f_error}) is larger than $0.01$\,PPT. 
    \item The target or reference star are detected by \textsc{SE} as two sources, as indicated by a deblending flag value of 2, due to the defocused PSF.
\end{itemize}

As a second filtering step, we further wish to remove images taken under non-photometric conditions, such as uneven clouds or haze across the field of view. During photometric conditions we expect the counts of the reference star to evolve smoothly with time (due to gradual airmass and seeing changes, for example). Departure from such smooth evolution can indicate non-photometric disturbances. We fit a piecewise second order polynomial to the reference star counts vs. time, such that each polynomial is fit to data from the same night, site, and two-hour time span, using only data points that are scattered less than a pre-determined threshold in 0.05-hour time bins. We then remove all images with a reference-star flux value that is farther than the pre-determined threshold from the polynomial. The pre-determined threshold is 15\,PPT for Campaigns 0, 1, and 2, and 250\,PPT for Campaign 4. The number of images retained after each filtering step is listed in Table~\ref{tab:analysis_num_images}.

We find an offset between the relative fluxes observed from different Las Cumbres sites. One possible explanation for this issue is the flat-field correction not being exact. To correct the offset, we subtract the median relative flux from the measured relative fluxes per site and night.

\begin{figure*}[]
   \includegraphics[width=\textwidth]{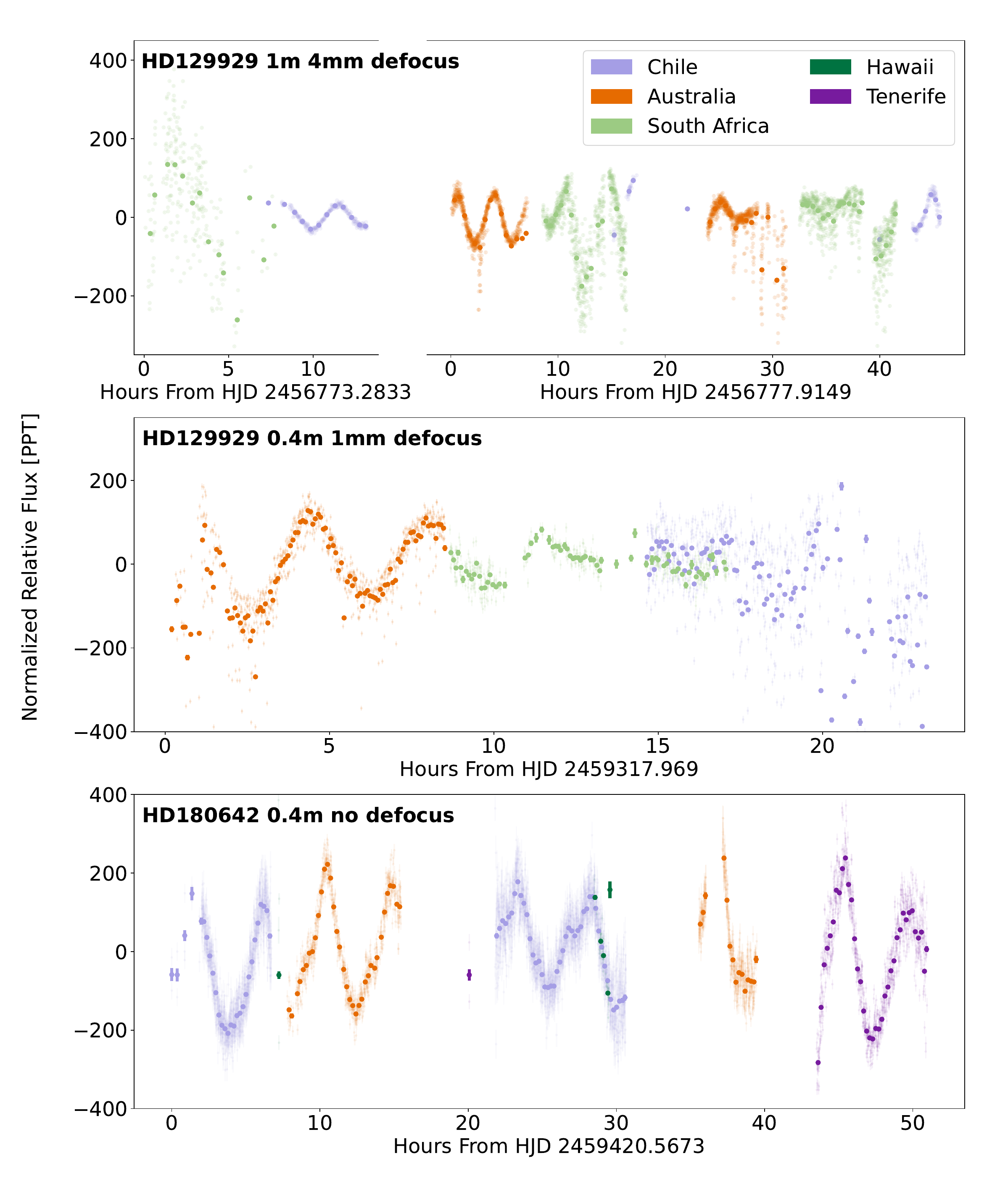}
\caption{Final light curves from Campaigns 0, 1, 2 and 4. The original data are plotted in semi-transparent points, and the data binned to $0.02$-day (top panel) $0.003$-day (middle panel), and $0.00833$-day (bottom panel) bins are plotted in opaque points. The oscillations (and varying amplitudes due to mode beating) can be clearly seen for most cases, with varying intrinsic scatter between sites and nights.}
\label{fig:final_lc}
\end{figure*}

\begin{figure}[]
\includegraphics[width=1\linewidth]{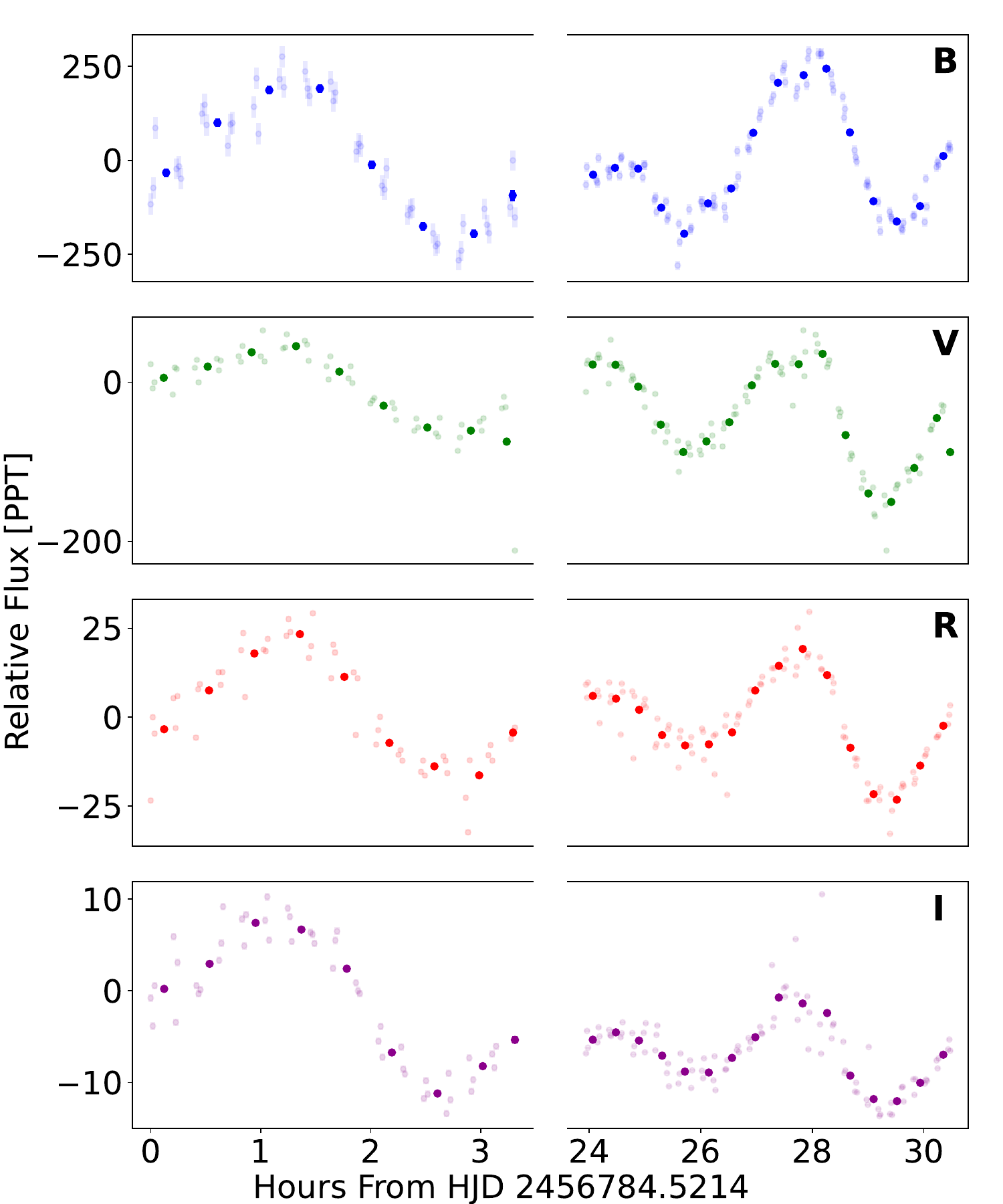}
\caption{Final light curves from Campaign 3 (the multi-band campaign). The oscillations are clearly seen, even when the camera switches filters after every third image. The original data are plotted in semi-transparent points, and the data binned to $0.0125$-day bins are plotted in opaque points.}
\label{fig:each_filter}
\end{figure}

Our final light curves are presented in Figures \ref{fig:final_lc} and \ref{fig:each_filter}. The stellar oscillations are clearly seen both in the single-band and multi-band campaigns, though some scatter remains, not removed by the filtering stages described above. The scattered points are not significantly correlated with detector temperature, sky brightness, photometric errors or a specific telescope or site. However, since the scatter is larger in the 1m data compared to the 0.4m data, we suspect that it is related to the non-trivial time-dependent PSF caused by defocusing. 

\begin{deluxetable}{lllll}
\tabletypesize{\footnotesize}
\tablecaption{Number of images retained after each filtering step.} \label{tab:analysis_num_images}
\tablehead{\colhead{Step} & \multicolumn{4}{c}{Campaign Number}\\
\colhead{} & \colhead{0} & \colhead{1} & \colhead{2} & \colhead{3}}
\startdata
Initial & 1146 & 4732 & 3636 & 5273\\
First Filtering & 1061 & 4403 & 2437 & 4511\\
Second Filtering & 517 & 3503 & 1861 & 4496\\
\enddata
\end{deluxetable}

\section{Light Curve Analysis}
\label{sec:lc_analysis}

These light curves serve as a test of a single 48-hour run. A full asteroseismic observation set will consist of several such runs. Therefore we do not expect to be able to recover all of the oscillation frequencies of our targets with the current observations. However, we can perform some ``sanity checks'' on -the data from our 48-hour runs (Campaigns 1 and 4).

\subsection{Model Fitting}
\label{sec:MCMC}
We fit the observations from Campaigns 1 and 4 to a sum of oscillations as described by: 

\begin{eqnarray} \label{eq:mcmc_model}
f(t) = a\sum_{i=0}^{N} a_i \cdot \sin(w_i(t-t_0) + \phi_i)\\ \nonumber
w_i = 2\pi f_i,
\end{eqnarray}
which is the same as Equation~\eqref{eq:artificial_signal} but with a global scaling factor $a$.

We use the \textsc{emcee} \citep{emcee} Python implementation of the Affine Invariant Markov Chain Monte Carlo (MCMC) Ensemble sampler \citep{MCMC} to fit Equation~\eqref{eq:mcmc_model} to our data, with $a$ and $t_0$ free parameters, and $a_i, \: f_i, \: \phi_i$ and $N$ the known amplitudes, frequencies, phases and total number of frequencies (respectively) of HD129929 \citep[from][]{Aerts2003} and HD180642 \citep[from][]{HD180642_2009}. We assume here that the targets continue to pulsate with the same frequencies, phases and relative amplitudes as when they were measured by those studies. This assumption is validated by the long-term monitoring of HD129929 by \cite{2004AA...415..241A}.

\begin{figure}[]
  \includegraphics[width=1\linewidth]{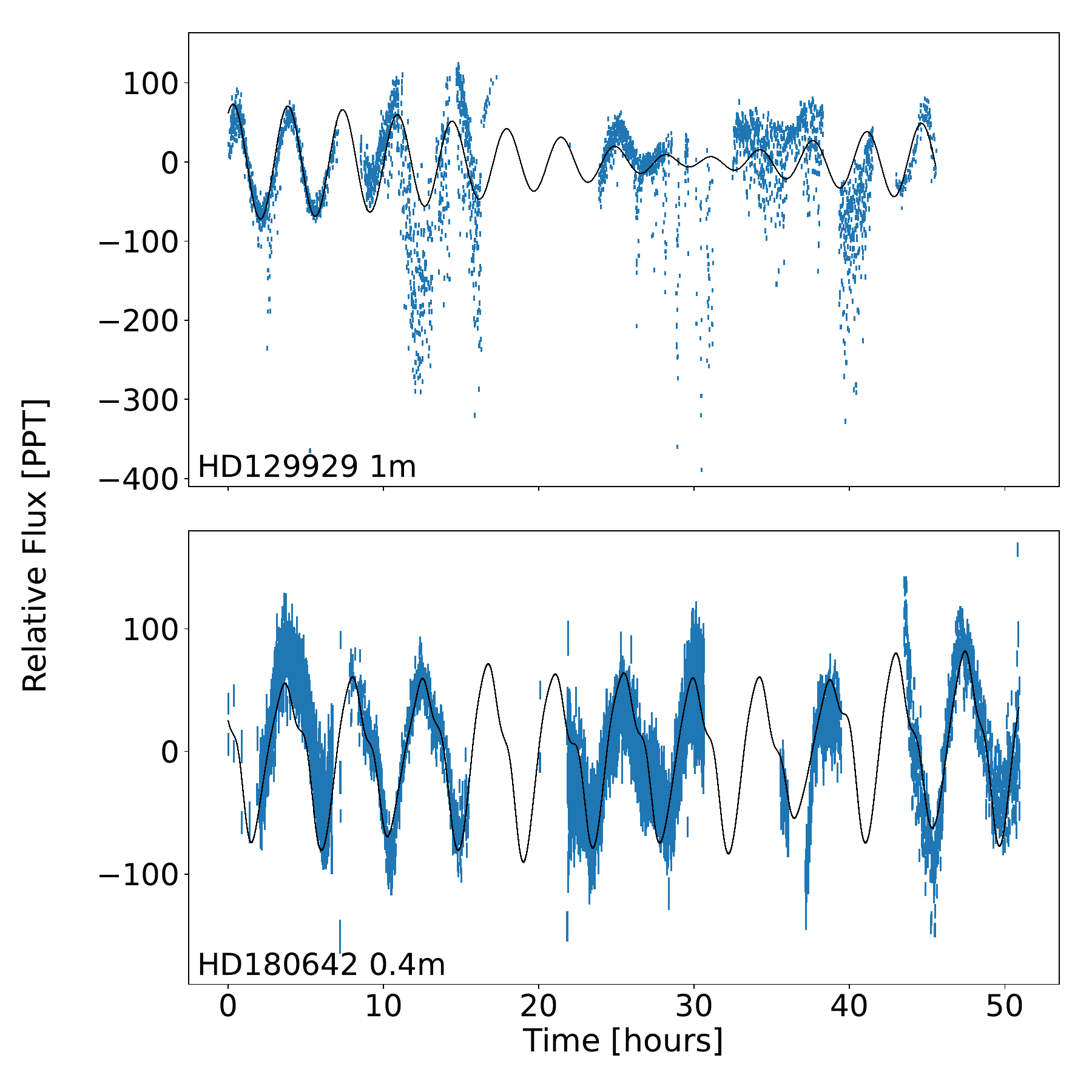}
\caption{MCMC fit to the observations from Campaign 1 (HD129929; top) and Campaign 2 (HD180642; bottom). 1000 random model light curves are plotted as drawn from the posteriors (black lines), but they appear as one thick line given that the posterior distributions are narrow. Time is in hours from the beginning of each observation (HJD 2456777.91486 for HD129929 and HJD 2459420.5672 for HD180642).}
\label{fig:MCMC_fit}
\end{figure}

The fit results are presented in Figure~\ref{fig:MCMC_fit} and their respective corner plots (generated with \textsc{corner.py}; \citealt{corner}) in Figure~\ref{fig:corner_plots} (Appendix \ref{appendix:MCMCfits}). We find good fits to the data, indicating that our observations are consistent with the expected pulsation parameters and with the assumption that these parameters remain stable over decades, as suggested by \cite{Aerts_book} for $\beta$~Cep stars.

\subsection{Periodograms} \label{sec:periodograms}
We calculate periodograms\footnote{All periodograms in this work are Lomb--Scargle \citep{Lomb1976, Scargle1982} periodograms, produced by the \texttt{to\_periodogram()} function of the \textsc{lightkurve} \textsc{Python} package \citep{2018ascl.soft12013L}.} for the data taken in Campaigns 1 and 4 and compare them to periodograms of artificial signals generated from Equation~\eqref{eq:artificial_signal} with the known parameters of our targets.

We sample the artificial signal (using the best fit value of $a$ from the MCMC runs described in Section \ref{sec:MCMC}) at the same time stamps as the actual observations, assuming typical measurement errors (3\,PPT\footnote{This is the typical error value for the 1m telescopes of the Las Cumbres Observatory, an order of magnitude higher error values gave similar results.}), and calculate its periodogram as well.
We repeat this step 160 times, each with a different time shift $t_0$ between 0 and 80 days in 0.5-day increments. 

\begin{figure}[]
   \includegraphics[width=1\linewidth]{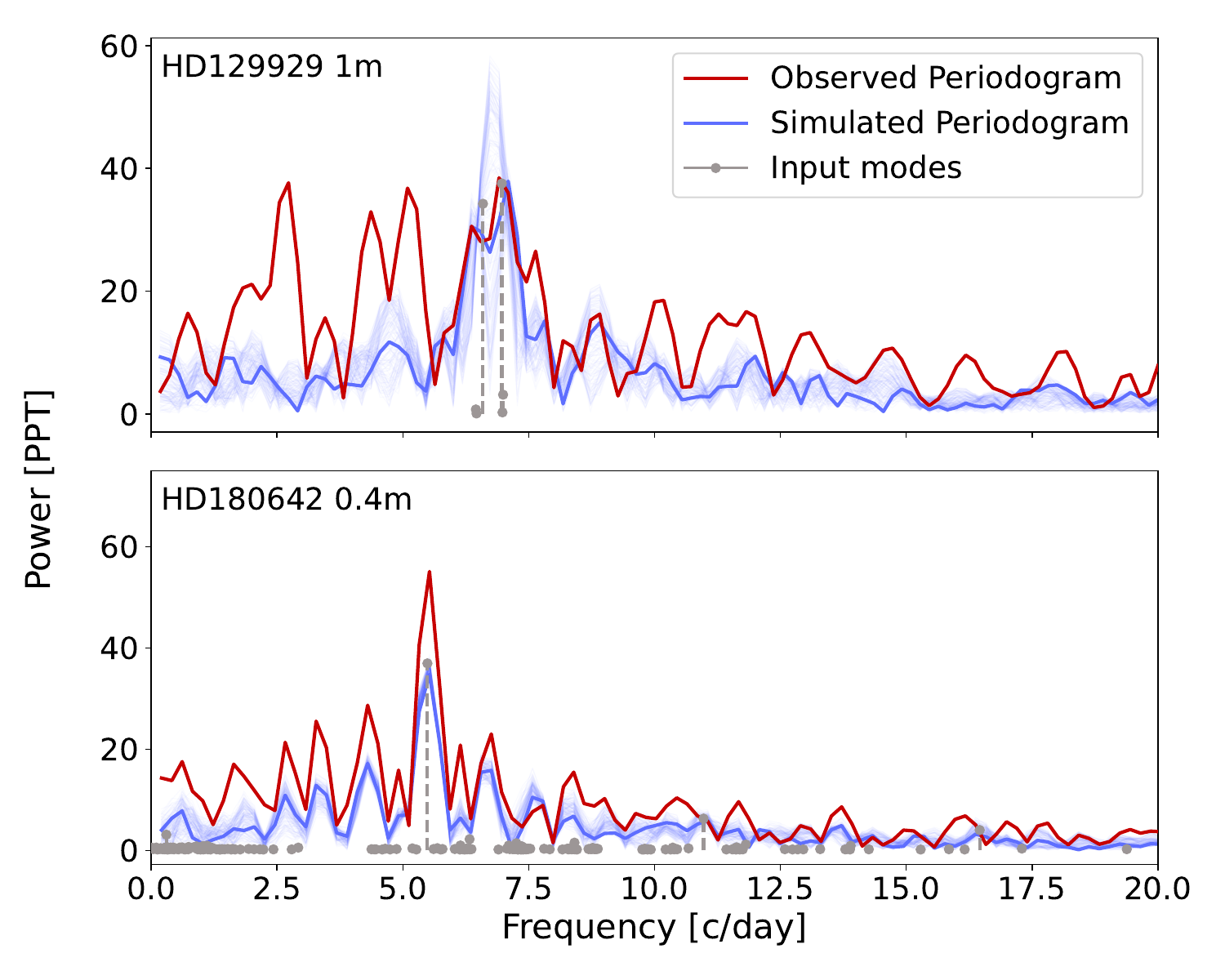}
\caption{Measured periodograms (red) for Campaign 1 (top) and 4 (bottom). Grey vertical dashed lines and circles denote the frequencies (at their respective amplitudes) of HD129929 and HD180642 (from \citealt{2004AA...415..241A} and \citealt{HD180642_2009}, respectively). The semi-transparent blue lines denote periodograms of 160 simulated signals with varying time offsets ($t_o$ in Equation~\eqref{eq:mcmc_model}). The opaque blue line denotes the periodogram of the simulated signal with the best fit time offset from the MCMC fit (see Section \ref{sec:MCMC}).}
\label{fig:periodograms}
\end{figure}

Our results are presented in Figure~\ref{fig:periodograms}. All the observed periodograms are roughly consistent with the artificial ones. The observed periodogram from Campaign 1 shows peaks at additional frequencies,  possibly due to the photometric scatter that our pipeline was unable to remove. 

\section{Discussion and Conclusions} \label{sec:conclusions}

We have shown that it is possible, in principle, to discern the oscillation frequencies of massive stars using a few weeks to months of observations, consisting of weekly high-cadence 48-hour runs. 

We then showed that it is possible, in practice, to schedule and obtain such $\sim$48-hour consecutive cross-site observations with the Las Cumbres Observatory network, both on its 1\,m and 0.4\,m telescopes. We further demonstrated that stellar oscillations of two different $\beta$~Cep stars can be distinguished by these observations, and that they match the expected oscillation patterns known for these stars. These oscillations are observable also when switching bands, allowing such data to be obtained in multiple bands, which is crucial for mode identification.

As the 0.4\,m telescopes require less defocusing than the 1\,m telescopes for the same stars, they are less likely to produce photometric artifacts that result from excessive defocusing. Thus, for bright enough targets, the 0.4\,m telescopes are more likely to provide light curves with lower scatter.

In principle, even the continuous observation runs don't have to be at such high cadence as was tested here. Sampling is only needed above the Nyquist limit. However, it remains to be tested how non-continuous (but still well-sampled) observations can be obtained through the Las Cumbres scheduler in practice.

This work serves as a proof-of-concept for using the Las Cumbres Observatory network to detect the oscillations of massive stars in multiple bands, in a way that can be scaled up to population studies of samples of objects. As such it forms the basis for the Global Asteroseismology Project which aims to obtain multi-band observations of $\sim$20 $\beta$ Cep stars.
Measuring and identifying the oscillation modes of a representative sample of massive stars will allow us to start to map the distribution of their internal properties from asteroseismic modeling \citep[e.g.][]{2009A&A...506..269B}. This will bring us one step closer to solving some of the largest puzzles regarding massive-star structure and evolution.

We further hope that this proof-of-concept will encourage performing continuous observations through the Las Cumbres Observatory network for other science cases as well, such as studies of other types of pulsating stars, binary systems, extra-solar planets, and even young supernovae, kilonovae, and other rapidly-evolving transients, all of which could benefit from continuous observations on few-day timescales.

~\\
We are grateful to the anonymous referee for their careful reading, knowledgeable comments, and constructive suggestions, all of which significantly improved this paper. We thank C.~Aerts for helping initiate the idea for this project, as well as for providing advice and guidance for its execution. We also thank L.~Bildsten, D.~Bowman, A.~Gilkis, J.~Goldberg, G.~Handler, D.~A.~Howell, M.~C.~Lam, D.~Maoz, T.~Mazeh, C.~McCully, M.~Michielsen, M.~G.~Pedersen, S.~Shahaf, R.~Street, and S.~Zucker for helpful discussions, useful advice, and insightful guidance. 
This work makes use of observations from the Las Cumbres Observatory global telescope network. We are grateful to the Las Cumbres Observatory staff for their help in understanding how to schedule these observations, and for their continued support of this unique facility.
NS is supported by a grant from the Prof. Amnon Pazy Research Foundation.
IA acknowledges support from the European Research Council (ERC) under the European Union’s Horizon 2020 research and innovation program (grant agreement number 852097), from the Israel Science Foundation (grant number 2752/19), from the United States - Israel Binational Science Foundation (BSF), and from the Israeli Council for Higher Education Alon Fellowship.

\vspace{5mm}
\facilities{Las Cumbres Observatory \citep[LCO;][]{LCO}}

\software{\textsc{astropy} \citep{astropy:2013, astropy:2018,                   
          astropy:2022},
          \textsc{LightKurve} \citep{2018ascl.soft12013L}, 
          \textsc{Astrometry.net} \citep{Astrometry}
          \textsc{Source Extractor} \citep{SExtractor}, 
          \textsc{Astro-SCRAPPY}, \citep{curtis_mccully_2018_1482019},
          \textsc{emcee} \citep{emcee}, corner.py \citep{corner}, IvS\footnote{\url{https://github.com/IvS-KULeuven/IvSPythonRepository}}
          }

\bibliography{sample631}{}
\bibliographystyle{aasjournal}

\appendix
\restartappendixnumbering

\section{Source Extractor Parameters} \label{appendix:SExtractor}

The parameters we changed from their default values in the configuration file of \textsc{SE} are presented in Table~\ref{tab:SE_config}. The rest of the parameters were set to the default values recommended by \textsc{SE}.
\startlongtable
\begin{deluxetable*}{lccccc}
   \tabletypesize{\footnotesize}
   \tablecaption{\textsc{SE} configuration file parameters (only those changes from their default values are shown).} \label{tab:SE_config}
   \tablehead{\colhead{Parameter} &\colhead{Units} & \multicolumn{4}{c}{Value per Campaign} \\ 
   \colhead{} & \colhead{} & \colhead{0,1,3 - SINISTRO} & \colhead{0,1,3 - SBIG} & \colhead{2} & \colhead{4} }
   \startdata
    DETECT\_MINAREA & (Pixels) & 10 & 10 & 10 & 10\\
    DETECT\_THRESH & & 5 & 5 & 5 & 5\\
    FILTER & & Y & Y & Y & Y\\
    FILTER\_NAME & & tophat\_5.0\_5$\times$4.conv & tophat\_5.0\_5$\times$4.conv & tophat\_5.0\_5$\times$4.conv & gauss\_4.0\_7$\times$7.conv \\
    DEBLEND\_NTHRESH & & 64 & 64 & 64 & 32\\
    DEBLEND\_MINCOUNT & & 1 & 1 & 1 & 0.005\\
    PHOT\_APERTURES & (Pixels) &  60 & 60 & 60 & 30\\
    SATURE\_LEVEL & (ADUs) & 60000 & 60000 & 60000 &60000\\
    PIXEL\_SCALE & (\arcsec/pixel) & 0.389 & 0.464 &  0.571 & 0.571\\
    CHECKIMAGE\_TYPE & & APERTURES & APERTURES & APERTURES & APERTURES
   \enddata
\tablecomments{See \cite{SExtractor} for more details about the parameters and their default values.}
\end{deluxetable*}

\section{Simulations} \label{appendix:simulations}

\subsection{The Prewhitening Process} \label{appendix:prewhitening}

\begin{figure*}[]
\centering
  \includegraphics[width=1\linewidth]{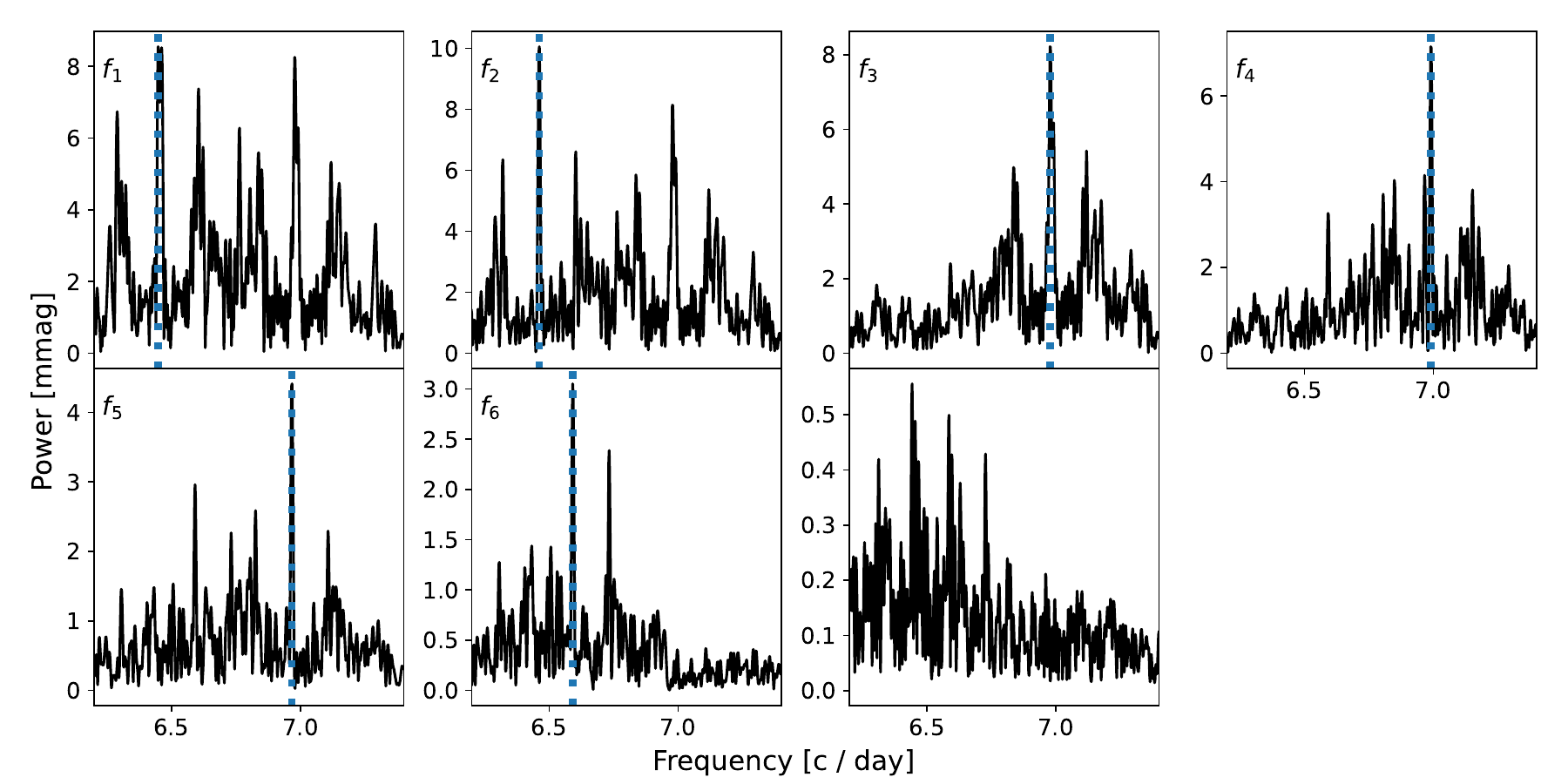} \caption{Lomb--Scargle periodograms after each prewhitening stage for the simulated light curve presented in Figure~\ref{fig:pre_observation_sim_HD129929}, with the highest amplitude frequency at that stage marked by a dashed blue line. The last panel (from left to right and top to bottom), is the residual periodogram after subtracting each frequency with the highest amplitudes in each step.}
  \label{fig:prewhitening_HD129929}
\end{figure*}

\begin{figure*}[]
\centering
  \includegraphics[width=1\linewidth]{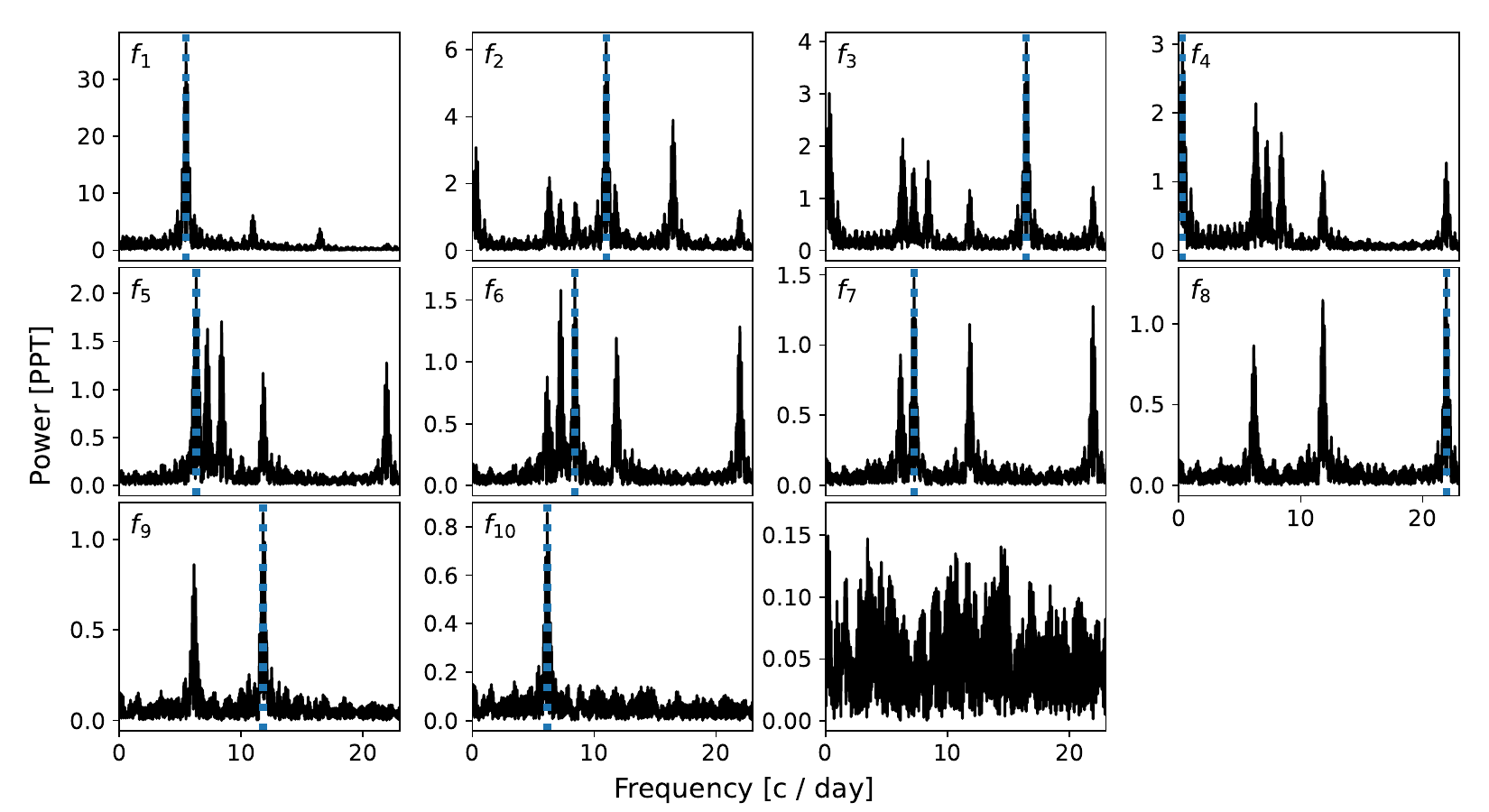}
  \caption{Same as Figure~\ref{fig:prewhitening_HD129929}, but for HD180642 and the simulated light curve presented in Figure \ref{fig:pre_observation_sim_HD180642}.}
  \label{fig:prewhitening_HD180642}
\end{figure*}

The residual periodograms for each prewhitening stage of the simulated data presented in Figure~\ref{fig:pre_observation_sim_HD129929} for HD129929 are shown in Figure~\ref{fig:prewhitening_HD129929}, and those for the simulated data presented in Figure~\ref{fig:pre_observation_sim_HD180642} for HD180642 are shown in Figure~\ref{fig:prewhitening_HD180642}.

\subsection{Monte Carlo Simulation Results} \label{appendix:MC_rand}

Results of the Monte Carlo simulations presented in Section~\ref{sec:monte_carlo_sim} for all six modes of HD129929, are given in Table \ref{tab:sim_1000_HD129929}.

\begin{deluxetable*}{c|D@{$\:\pm$}DD@{$\:\pm$}DD@{$\:\pm$}DD@{$\:\pm$}D}
\tablecaption{The average $\pm$ standard deviation of the differences between recovered frequencies and amplitudes and the ones input to the simulations.} \label{tab:sim_1000_HD129929}
\tablehead{\colhead{Y (hrs)} & \multicolumn2c{12} & \multicolumn2c{} & \multicolumn2c{24} & \multicolumn2c{} & \multicolumn2c{36} & \multicolumn2c{} & \multicolumn2c{48} & \multicolumn2c{} \\ \hline \colhead{} & \multicolumn{16}{c}{$\Delta f \: (c/day)$}}
\decimals
\startdata
$f_1$ & 0.0478 & 0.1458 & 0.0105 & 0.0676 & 0.0000 & 0.0061 & 0.0001 & 0.0041 \\
$f_2$ & -0.1291 & 0.2253 & -0.0222 & 0.1032 & -0.0022 & 0.0333 & -0.0005 & 0.0167 \\
$f_3$ & 0.1606 & 0.2524 & 0.0256 & 0.1121 & 0.0049 & 0.0506 & 0.0010 & 0.0239 \\
$f_4$ & -0.0997 & 0.2364 & -0.0165 & 0.0928 & -0.0030 & 0.0388 & -0.0008 & 0.0212 \\
$f_5$ & 0.2168 & 0.3012 & 0.2570 & 0.1801 & 0.2591 & 0.1737 & 0.2176 & 0.1852 \\
$f_6$ & -0.2297 & 0.3793 & -0.2597 & 0.1826 & -0.2587 & 0.1748 & -0.2176 & 0.1852 \\ \hline
\multicolumn{16}{c}{$\:\:\:\:\:\:\:\:\:\:\:\:\:\:\:\:\:\:\:\:\:\:\:\:\:\:\:\:\:\:\:\:\Delta A \: (mmag)$}\\ \hline
$f_1$ & -0.0918 & 1.2383 & -0.1504 & 0.5061  & -0.0571 & 0.2231 & -0.0367 & 0.1177 \\
$f_2$ & 0.1040 & 1.1461 & -0.1805 & 0.4841 & -0.0540 & 0.1825 & -0.0217 & 0.1040 \\
$f_3$ & -0.7035 & 1.1761 & -0.2119 & 0.5641 & -0.0816 & 0.3057 & -0.0576 & 0.2118 \\
$f_4$ & -0.7350 & 0.9958 & -0.2106 & 0.3958 & -0.0845 & 0.1831 & -0.0567 & 0.1072 \\
$f_5$ & -0.2738 & 0.6537 & -0.1678 & 0.3699 & -0.1812 & 0.1990 & -0.1597 & 0.1191 \\
$f_6$ & -1.0334 & 0.6982 & -0.6206 & 0.5313 & -0.4063 & 0.3537 & -0.2891 & 0.2383 \\
\enddata
\end{deluxetable*}

\section{MCMC Fits} \label{appendix:MCMCfits}

\begin{figure*}[]
\centering
  \plottwo{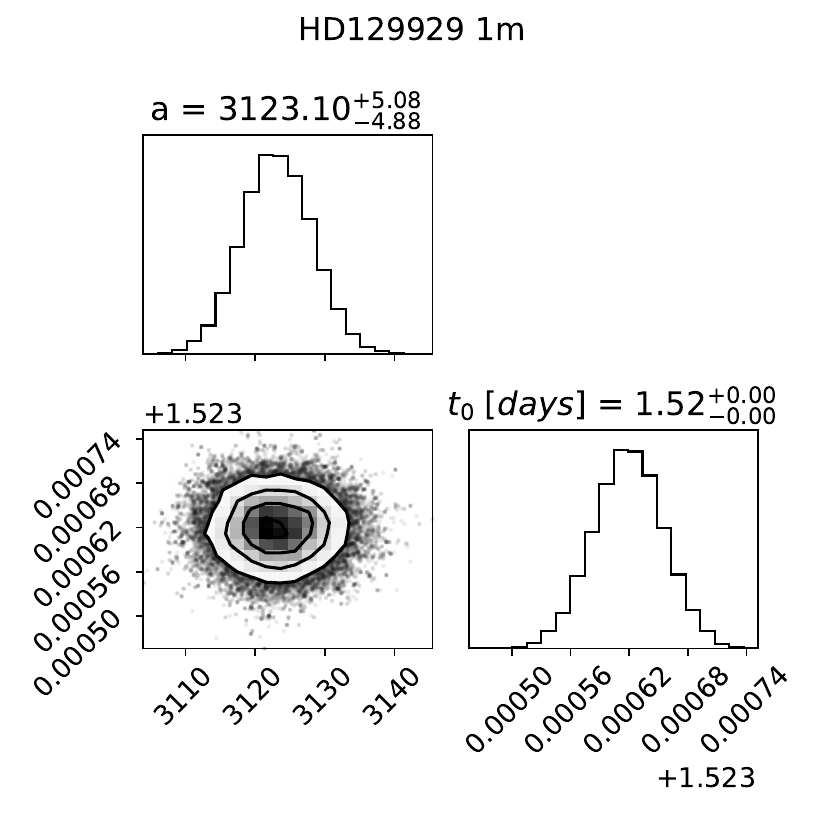}{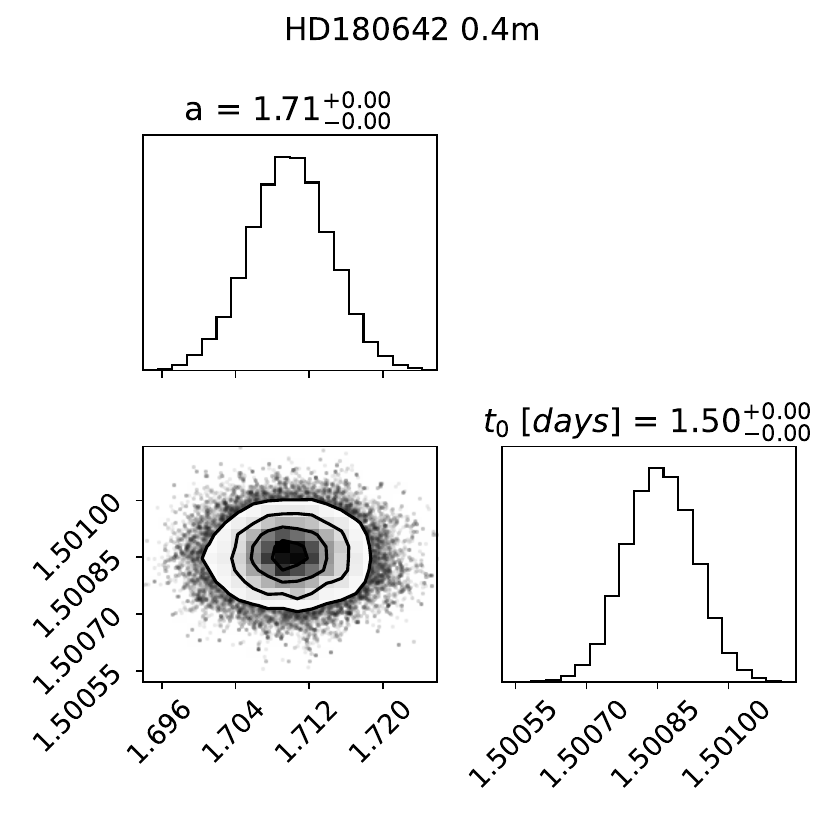}
\caption{Corner plots of the two MCMC fits presented in Figure~\ref{fig:MCMC_fit} for Campaigns 1 (left) and 4 (right).}
\label{fig:corner_plots}
\end{figure*}

The corner plots from the MCMC fits presented in Figure~\ref{fig:MCMC_fit} are shown in Figure~\ref{fig:corner_plots}. 

\end{document}